\documentclass{aa}
\usepackage{graphicx}
\def\funits{$\rm erg~cm^{-2}~s^{-1}$}
\def\cunits{$\rm cm^{-2}$}

\def\xmm{{\it XMM-Newton~}}

\def\rosat{{\it ROSAT~}}
\def\asca{{\it ASCA~}}

\begin{document}
 \title{On the XMM-Newton spectra of soft X-ray selected QSOs 
	}

%   \subtitle{ }

  \titlerunning{XMM-Newton QSO spectra}
    \authorrunning{A. Akylas et al.}

   \author{A. Akylas\inst{1,2}
	  I. Georgantopoulos\inst{1}
          X. Barcons \inst{3}
          }

   \offprints{I. Georgantopoulos, \email{ig@astro.noa.gr}}

   \institute{Institute of Astronomy \& Astrophysics,
              National Observatory of Athens, 
 	      Palaia Penteli, 15236, Athens, Greece \\
        \and 
	 Physics Department, University of Athens, Panepistimiopolis, 
	 Zografos, 15783, Athens, Greece \\
         \and
         Istituto de Fisica de Cantabria, (CSIC-UC), 39905, Santander, Spain  
             }

   \date{Received ; accepted }

   \abstract{
 We study the \xmm spectra of a sample 
 of 32 soft X-ray selected QSOs.
 Our goal is to check, using the spectra
 of moderate redshift ($z\sim1.5$), 
 faint ($\rm f_{0.2-8 keV}>5\times 10^{-15}$ \funits) 
 broad-line  QSOs, previous 
 claims for either significant intrinsic absorption 
 or spectral hardening at high energies.  
 We derive hardness ratios  
  for all sources and furthermore 
 we perform spectral fits for the 11 brighter sources.
 The majority of sources have steep spectra $\Gamma>1.9$.   
 We find a few QSOs with large amounts 
 of intrinsic absorption, as high as $\rm N_H\sim10^{23}$ \cunits.
 We find no strong evidence for spectral hardening above 2 keV.  
 The coadded QSO spectrum      
 is well described by a single power-law 
 with photon index of $\sim 1.9$,
 demonstrating that, on average, any effects of absorption 
 are not important.
 This suggests that the discrepancy between the 
 X-ray background and the (soft X-ray selected) QSO 
 spectrum holds well at the faint fluxes probed here.  
	}

   \maketitle
%
%________________________________________________________________

\section{Introduction}

In the last decade  there has been a great 
progress in understanding  the X-ray spectral 
 properties of QSOs 
 (see Mushotzky et al. 1993 for a review). 
{\it GINGA}  (Lawson \& Turner 1997) and \asca (Reeves \& Turner 2000) 
observations of nearby, bright QSOs (typically  with flux 
$>10^{-12}~\rm erg~s^{-1}~cm^{-2}$) in the 2-10 keV band 
have shown a
power-law spectrum ($\Gamma\sim 1.9$) with no 
evidence  for absorption  above the Galactic value. 
At softer energies  (0.1-2 keV) \rosat observations show 
a much steeper spectrum ($\Gamma \sim 2.5$, Laor et al. 1997). 
 This steepening 
is attributed to either an additional soft component 
at energies $<$0.5 keV or simply to calibration 
uncertainties between  different instruments
(Iwasawa et al. 1999).
In any case, the bright QSO spectra  in the 
0.5-10 keV band are much steeper 
than the spectrum of the X-ray background (XRB) which has 
$\Gamma = 1.4$ (Gendreau et al 1995,
Miyaji et al 1998, Vecchi  et al. 1999). 
This spectral mismatch known as the spectral paradox  
suggested that QSOs cannot produce the bulk of the 
X-ray background. 

However, the spectrum of a broad-line  
$\rm L_\star$ QSO (those which contribute 
 a significant fraction of the XRB) 
 at large distances
($z\sim 1.5$) is largely  unknown. 
Schartel et al. (1996), Blair et al. (2000) studied 
the co-added spectra of faint  
distant QSOs with \rosat. They find an average spectrum of 
 $\Gamma \sim 2.1-2.6$ flattening with increasing redshift.
Pappa et al. (2001) studied the 
{\it ROSAT/ASCA}  spectrum of 21 hard X-ray selected QSO. 
They find evidence for spectral 
curvature  in the sense that the 
2-10 keV \asca  spectrum  has
$\Gamma\sim1.5$ while  their \rosat
spectrum  is steep $\Gamma \sim 2.2 $. More recently, 
Barcons et al. (2002) found evidence for 
such a spectral curvature in the average QSO hardness 
ratio in their \xmm SSC Medium Sensitivity survey.
 They find $\Gamma\sim2$ and $\Gamma\sim1.6$ 
 at soft (0.5-4.5 keV) and hard (2-10 keV) energies respectively.  
Moreover, it becomes  now evident that some QSOs present 
high amounts of obscuration (e.g. Fiore et al. 1999).
The above results  bear great 
significance as they may suggest 
that QSOs can contribute a larger
fraction  of the X-ray background  than that
usually predicted  by the standard synthesis models
 (e.g. Comastri et al. 1995). 

 In this paper we exploit 
 the large effective  area and the 
 extended  passband of  \xmm
 in order to analyze the  spectral 
 properties of   
 faint $\rm f_{0.2-8 keV}> 5\times 10^{-15}$ \funits 
 ($ \Gamma=2$) Broad-Line QSOs. 
 In particular, we analyze the  spectral properties  
 of 32 soft X-ray selected,  
 broad-line (type-I), QSOs in the Lockman Hole (Schmidt
 et al. 1998, Lehmann et al. 2001).
  These QSOs contribute 
  a significant fraction ($\sim50$\% in the 0.9-2 keV band)  
  of the X-ray background intensity
 (Hasinger et al 1993).     
 Results on the \xmm 
 properties of all sources in the  
 Lockman Hole have been presented by 
 Hasinger et al. (2001) and Mainieri et 
 al. (2002). Here instead,  
 we put emphasis on testing  for
 the presence of absorption or  
 spectral flattening at high energies    
 in the individual or co-added QSO spectra 
 and the implications for the X-ray 
 background.

\section{The X-ray Data}

We use \xmm data from the Lockman Hole Observation, 
centered on the sky position RA 10:52:43, DEC +57:28:48 (J2000). 
In particular we use data from three \xmm observations, 
(revolutions 70, 73 and 74) with total 
 exposure time (on-time) of  $\sim$ 120 ks.
 The astrometry offsets between the 
 3 revolutions above are within only a couple of arcsec 
 and therefore they are not taken into account. 
 We analyse the pipeline products using the \xmm Science Analysis 
System ({\sc sas} v5.3). We deal only with data from the EPIC-PN 
camera because its sensitivity is significantly 
greater than that of MOS (Str\"uder et al 2001). 
The EPIC-PN camera was operated in the standard full-frame mode.
The pixel size corresponds to $\sim$4.1 arcsec.
For on-axis  point sources, a circle  of radius $\sim$ 32 arcsec 
 includes  90 \% of the 1.5 keV photons 
 and 85 \% of the 5 keV photons.
 The Point--Spread--Function (PSF)  
 does not strongly depend on the off-axis angle.  
 For sources at 7 arcmin off-axis 
 the above radius encircles almost the same percentage 
 of soft and hard photons. 
 The vignetting correction is  
 $\sim$ 28 \% for 1.5 keV  photons and $\sim$31 \% 
 for 5 keV photons at 7 arcmin off-axis.  
The thin filter is used in all three observations. 
We construct the event file using 
single and double events (patterns 0--4). 

 The signal-to-noise ratio of the sources  
 becomes low at high energies  
 due to the presence of a strong  Cu-Ka line at 8.1 keV.  
 Therefore we have excluded from our analysis photons with 
 energies $>$8 keV. 
 A substantial fraction of the observations were also
 affected by high particle background 
 with count rate up to several hundred per second, compared to 
 a quiescent count rate of several counts per second.  
 We locate flares by analyzing the full field--of--view (FOV) light curves. 
 We reject all the time intervals with count rates, 
 in the 0.2-8 keV band, 
 higher than 10 cts/s. 
 The remaining good time intervals give an exposure of $\sim$ 50 ks. 

 We extract four images in the following bands: 
 0.2-0.5, 0.5-2, 2-4.5 and 4.5-8 keV. 
 Exposure maps, which account for vignetting, CCD gaps,
 bad columns, and bad pixels, are constructed for each band. 
 We also extract background maps for each energy band. 
 We use the {\sc eboxdetect} task 
 to search for sources in the four images simultaneously. 
 We use a total detection likelihood of 18. This corresponds to
 less than  one spurious detection per image. 
 The same routine was used to derive the hardness ratios.

 We obtain the  
 optical identifications of the detected sources 
 from  the 
 \rosat Ultra Deep Survey Optical Identification 
 catalogue (Lehmann et al 2001).
 We select all sources classified as type I QSOs 
 by  Lehmann et al (2001) i.e.  these with 
 optical spectroscopic class a, b or c.
 There are 44 QSOs within the \xmm FOV:
 3 fall on CCD gaps or bad columns, 2 are not detected 
 while 7 are not considered for spectral analysis as they 
 are faint.  
 The resulting sample contains 32 QSOs.   
 Four of these are associated with radio sources 
  in the sample of de Ruiter et al (1997).
 Hereafter, we refer to these as the ``radio-loud'' QSOs,
 regardless of whether these follow 
 the criterion 
 $\rm L(5GHz)>2.5\times 10^{24}h^{-2}_{100}  W Hz^{-1}$
 (Kellermann et al. 1989).   
 The sample covers  a redshift range  
from 0.2 to 3.4, with a mean of z$\sim1.5$. 
 The lower and upper fluxes are  
$ 5.3\times 10^{-15}$ and  $6.5\times 10^{-13}$ 
$\rm ergs~s^{-1}~cm^{-2}$ respectively in the 
 0.2-8 keV band, assuming a power-law model with $\Gamma=2$.

\section{Spectral Analysis}

 We determine the spectral properties  
 using both spectral fitting and the hardness ratios. 
 We extract spectral files for the 11 brighter 
 ($\rm f_{0.2-8 keV}>4\times10^{-14}$ \funits) type-I QSO
 in the case where the signal to noise ratio is relatively high.  
 Moreover, we calculate the Hardness Ratios (HR)
 for all sources. In Table \ref{counts} we give the 
extracted source and background counts in the 0.2-8 keV energy band 
for the 32 QSOs as obtained from the {\sc emldetect} task 
in the {\sc sas} analysis software.

\begin{table*}
\begin{center}
\caption{The Hardness Ratios}
\label{counts}
\begin{tabular}{ccccccc}

\hline
Name & No & z & source counts & error & background counts \\
     &    &   &  $\times 10^{2}$ &  $\times 10^{2}$ & $\times 10^{2}$  \\
\hline
\hline 

RXJ105125.4+573050 & 1 & 3.40 &  1.52 &  0.20 &  0.54  \\     
RXJ105144.8+572808 & 2 & 3.40 &  1.64 &  0.19 &  0.53 \\      
RXJ105154.4+573438 & 3 & 0.87 &  6.62 &  0.34 &  0.61 \\      
RXJ105213.3+573222 & 4 & 1.87 &  1.86 &  0.20 &  0.59 \\      
RXJ105224.7+573010 & 5 & 1.00 &  1.60 &  0.18 &  0.54 \\      
RXJ105228.4+573104 & 6 & 0.93 &  1.17 &  0.18 &  0.61  \\     
RXJ105230.3+573914 & 7 & 1.44 &  5.29 &  0.32 &  0.62  \\        
RXJ105239.7+572432 & $^\star$8 &1.11&38.7 &0.77 &0.45  \\        
RXJ105243.1+571544 & 9 & 2.14 &  3.00 &  0.26 &  0.52  \\        
RXJ105245.7+573748 & 10 & 1.68&  0.75 &  0.17 &  0.69  \\        
RXJ105247.9+572116 & 11& 0.46 &  10.3 &  0.40 &  0.39  \\        
RXJ105254.3+572343 & 12 & 0.76&  4.88 &  0.28 &  0.35   \\       
RXJ105257.1+572507 & 13 & 1.52&  4.31 &  0.27 &  0.41  \\        
RXJ105259.2+573031 & 14 & 1.67&  3.73 &  0.28 &  0.81  \\        
RXJ105302.6+573759 & 15 & 1.88&  3.46 &  0.27 &  0.71  \\        
RXJ105303.9+572925 & 16 & 0.78&  3.26 &  0.25 &  0.75  \\        
RXJ105306.2+573426 & 17 & 2.94&  1.45 &  0.20 &  0.81  \\        
RXJ105307.2+571506 & $^\star$18 &2.41 &1.61&0.21 & 0.48    \\     
RXJ105309.4+572822 & 19 & 1.56&  3.83 &  0.27 &  0.71  \\       
RXJ105312.5+573425 & 20 & 1.20&  2.70 &  0.24 &  0.81  \\       
RXJ105312.4+572507 & 21 & 0.96&  1.16 &  0.17 &  0.60   \\       
RXJ105316.8+573552 & $^\star$22 &1.2 & 36.2 & 0.76 &  0.75    \\      
RXJ105322.2+572852 & 23 & 2.30&  1.38 &  0.19 &  0.76    \\       
RXJ105324.7+572819 & 24 & 1.51&  1.17 &  0.19 &  0.76   \\      
RXJ105329.2+572104 & 25 & 1.14&  1.57 &  0.19 &  0.46   \\      
RXJ105331.8+572454 & 26 & 1.95&  10.4 &  0.41 &  0.66   \\      
RXJ105335.1+572542 & 27 & 0.78&  26.3 &  0.65 &  0.69   \\     
RXJ105339.7+573105 & 28 & 0.58&  20.7 &  0.57 &  0.70   \\  
RXJ105344.9+572841 & 29 & 1.81&  5.15 &  0.30 &  0.74   \\  
RXJ105350.3+572710 & 30 & 1.70&  4.49 &  0.29 &  0.67  \\ 
RXJ105358.5+572925 & 31 & 1.84&  0.95 &  0.17 &  0.64  \\
RXJ105421.1+572545 & $^\star$32& 0.20 & 56.1&1.02&  0.46  \\

\end{tabular}
\end{center}
$^\star$ Radio Loud QSO
\end{table*}      

\subsection{The Spectra of the brighter sources}

\begin{figure}
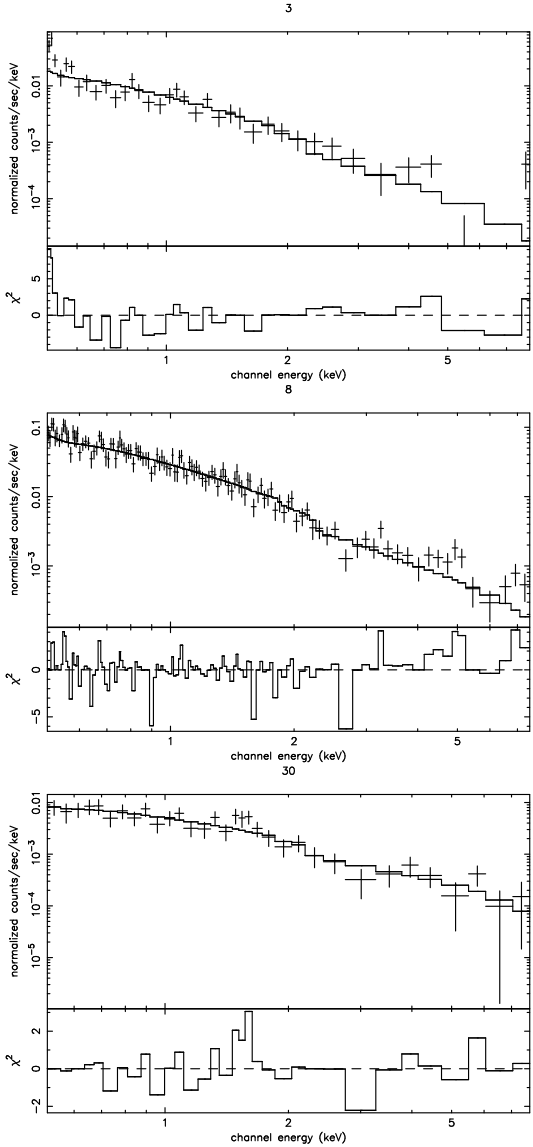

 \centering
\rotatebox{270}{\includegraphics[height=7.0cm]{h3736_fig1a.ps}}   
\rotatebox{270}{\includegraphics[height=7.0cm]{h3736_fig1b.ps}}
\rotatebox{270}{\includegraphics[height=7.0cm]{h3736_fig1c.ps}}   
 \caption{Power-law fits and $\chi^2$ residuals 
 to the spectra of the sources \#3, 8 and 30}
 \label{fig_spectra}
 \end{figure}

We derive the spectral files of the 
11 brighter sources by   
using an extraction radius of 32 arcsec ($\sim$ 8 pixel). 
We extract the background spectrum 
from parts of the image which do not 
 contain any obvious sources.  
 We use the response matrix 
$\rm epn\_ff20\_sdY9.rmf$ provided by 
 the \xmm calibration page (http://xmm.vilspa.esa.es) 
 which takes into account both the   
 single and double events. 
 We finally create auxiliary files at every off-axis angle
  using the ARFGEN task. 
 The source spectra were grouped to give a minimum number
 of 15 counts per bin so that Gaussian statistics apply. 
 The spectrum of all the sources was fitted in XSPEC v11.
 The errors quoted in the spectral analysis with XSPEC 
 correspond to the 90 \% confidence level. 
 We fit the spectra in the 0.5-8 keV band,
 using an absorbed power--law model.
 However, only in one case (source \#32)  
 we detect an equivalent neutral hydrogen column density, 
 $\rm N_H$, significantly greater than the Galactic  
 column of $6\times 10^{19}~ \rm cm^{-2}$ (Dickey \& Lockman 1990).  
 In Table \ref{spectra} we present the spectral fit results.
 We obtain good fits in most cases. Source \#3 presents a 
 bad $\chi^2$ suggesting a complex spectrum 
 (see Fig. \ref{fig_spectra}).
 In the four brighter cases (sources \# 8, 22, 27, 32) 
 we attempt to fit the data using a more complicated model 
 consisting of two power-law components.    
 We obtain a statistically significant improvement 
 only in the case of source \#8;  
 $\Delta\chi^2\approx9.5$ for an additional parameter
 which is significant at over the 99 \% confidence level. 
 The high energy power-law is fixed at $\Gamma=1.9$ 
 while the soft power-law is $\Gamma=3.45^{+0.32}_{-0.28}$. 
 The {\it single} power-law spectrum of this source is 
 also plotted in Fig. \ref{fig_spectra}.

\begin{table*}
\begin{center}
\caption{ Spectral fitting results for the  11 brighter sources}
\label{spectra} 
\begin{tabular}{cccccccc}

\hline
Name & No & z & $N_H$ & $\Gamma$ & $\chi^2$/d.o.f    \\
 & &   &  $ \rm \times10^{22} cm^{-2}$ &        &    \\
\hline
\hline 
RXJ105154.4+573438 & 3  & 0.87 & $0.^{+0.03}$ & $2.41^{+0.26}_{-0.24}$ & 60.1/34 & \\
RXJ105230.3+573914 & 7  & 1.44 &	 $0.^{+0.08}$           & $2.06^{+0.53}_{-0.22}$ &  14.5/27 & \\
RXJ105239.7+572432 & 8$^\star$  & 1.11 &$0.^{+0.01}$	    & $2.27^{+0.08}_{-0.08}$ & 126.6/119 & \\
RXJ105247.9+572116 & 11  & 0.46 &   $0.^{+0.04}$          & $2.39^{+0.26}_{-0.19}$ & 32.3/40 & \\
RXJ105316.8+573552 & 22$^\star$  &1.20 &   $0.^{+0.01}$   & $1.87^{+0.05}_{-0.07}$ & 120.4/136 & \\
RXJ105331.8+572454 & 26  & 1.95 &	$0.03^{+0.05}_{-0.03}$  & $2.17^{+0.30}_{-0.25}$ & 51.9/53 & \\
RXJ105335.1+572542 & 27  & 0.78 &$0.04^{+0.03}_{-0.03}$      & $2.15^{+0.15}_{-0.15}$ & 96.7/107 & \\
RXJ105339.7+573105 & 28  & 0.58 &       $0.^{+0.02}$  & $2.32^{+0.14}_{-0.09}$ & 85.0/73 & \\
RXJ105344.9+572841 & 29 & 1.81 &$0.^{+0.07}$     & $1.93^{+0.42}_{-0.22}$ & 24.4/29 & \\
RXJ105350.3+572710 & 30 & 1.70 &$0.^{+0.08}$     & $1.67^{+0.42}_{-0.18}$ & 21.8/30 & \\
RXJ105421.1+572545 & 32$^\star$  &0.20 &  $0.15^{+0.03}_{-0.02}$  & $1.87^{+0.09}_{-0.06}$ &261.0/257& \\
\hline
\end{tabular}
\end{center}
$^\star$ Radio Loud QSO
\end{table*}

\subsection{The Hardness Ratios}

We calculate three hardness ratios using 
 the source counts in the following energy bands 
  S1=0.2-0.5 keV, S2=0.5-2 keV, M=2-4.5 keV and H=4.5-8 keV.
  The hardness ratios are defined as 
  HR1=(S2-S1)/(S2+S1), HR2=(M-S2)/(M+S2), HR3=(H-M)/(H+M).
  The hardness ratios are estimated using the 
 {\sc emldetect} task in {\sc sas},  
  which corrects 
 both for vignetting and for the light falling 
 outside the detection circle (PSF correction).   
 The hardness ratios are given in Table \ref{master}
 together with their $1\sigma$ uncertainties.
 The uncertainties are estimated using Poisson 
 statistics and error propagation.    
 Given the energy range of 
 each  HR, HR1 is more sensitive    
 to the presence of a soft excess component or  
 even low amounts of
 column density ($\> 10^{20}~\rm cm^{-2} $).
 HR2 is affected by the presence of column densities   
 $\sim10^{21}~\rm cm^{-2} $ while 
 HR3 values could be sensitive to either   
 a reflection component or to even 
 larger columns (ie $> 10^{22}~\rm cm^{-2} $).

\begin{table*}
\begin{center}
\caption{The Source List}
\label{master}
\begin{tabular}{ccccccc}

\hline
Name & No & z & log(0.2-8 keV flux)$^1$  & HR1 & HR2 & HR3  \\
     &    &   & $\rm ergs~s^{-1}~cm^{-2}$ &   & &   \\  
\hline
\hline 

RXJ105125.4+573050 & 1 		& 3.40 	& -13.89 $\pm$0.058	& 0.513             $\pm0.139$ 	& -0.724  $\pm0.105$  &-0.811  $\pm0.654$ \\ 
RXJ105144.8+572808 & 2 		& 3.40 	& -13.96 $\pm$0.052	& 0.816  	    $\pm0.134$	& -0.645  $\pm0.094$  &-0.271  $\pm0.276$ \\
RXJ105154.4+573438 & 3 		& 0.87 	& -13.33 $\pm$0.022	& 0.223  	    $\pm0.054$	& -0.702  $\pm0.046$  &-0.689  $\pm0.167$ \\
RXJ105213.3+573222 & 4 		& 1.87 	& -14.00 $\pm$0.048	& 0.450  	    $\pm0.116$	& -0.783  $\pm0.084$  &-0.275  $\pm0.401$ \\
RXJ105224.7+573010 & 5 		& 1.00 	& -14.14 $\pm$0.050	& 0.357  	    $\pm0.120$	& -0.729  $\pm0.098$  &-0.527  $\pm0.421$ \\
RXJ105228.4+573104 & 6 		& 0.93 	& -14.26 $\pm$ 0.066	& 0.162  	    $\pm0.185$	& -0.505  $\pm0.165$  & 0.016  $\pm0.288$ \\
RXJ105230.3+573914 & 7 		& 1.44 	& -13.35 $\pm$0.026	& 0.351  	    $\pm0.063$	& -0.724  $\pm0.051$  &-0.386  $\pm0.190$ \\
RXJ105239.7+572432 & $^\star$8	& 1.11 	& -12.73 $\pm$0.008	&-0.014             $\pm0.021$	& -0.738  $\pm0.019$  &-0.421  $\pm0.064$ \\
RXJ105243.1+571544 & 9 		& 2.14 	& -13.51 $\pm$0.038	& 0.248  	    $\pm0.089$	& -0.759  $\pm0.075$  &-0.557  $\pm0.420$ \\
RXJ105245.7+573748 & 10 	& 1.68	& -14.27 $\pm$0.100	& 0.436 	    $\pm0.257$	& -0.891  $\pm0.167$  & 0.006  $\pm1.286$ \\
RXJ105247.9+572116 & 11		& 0.46 	& -13.20 $\pm$  0.017	&-0.113 	    $\pm0.040$  & -0.784  $\pm0.038$  &-0.700  $\pm0.185$ \\
RXJ105254.3+572343 & 12 	& 0.76	& -13.60 $\pm$0.025	& 0.321  	    $\pm0.061$	& -0.607  $\pm0.058$  &-0.392  $\pm0.144$ \\
RXJ105257.1+572507 & 13 	& 1.52	& -13.69 $\pm$0.027	& 0.089  	    $\pm0.064$	& -0.852  $\pm0.053$  &-0.660  $\pm0.432$ \\
RXJ105259.2+573031 & 14 	& 1.67	& -13.77 $\pm$0.032	& 0.413  	    $\pm0.087$	& -0.550  $\pm0.070$  &-0.405  $\pm0.155$ \\
RXJ105302.6+573759 & 15 	& 1.88	& -13.57 $\pm$0.033	& 0.063  	    $\pm0.083$	& -0.625  $\pm0.079$  &-0.542  $\pm0.243$ \\
RXJ105303.9+572925 & 16 	& 0.78	& -13.83 $\pm$0.033	& 0.225  	    $\pm0.081$	& -0.727  $\pm0.066$  &-1.000  $\pm0.264$ \\
RXJ105306.2+573426 & 17 	& 2.94  & -14.01 $\pm$0.061	& 0.257  	    $\pm0.157$	& -0.651  $\pm0.125$  &-0.519  $\pm0.433$ \\
RXJ105307.2+571506 & $^\star$18 & 2.41  & -13.73 $\pm$0.059	& 0.332     	    $\pm0.143$	& -0.444  $\pm0.130$  &-1.00 \ $\pm0.408$ \\ 
RXJ105309.4+572822 & 19 	& 1.56 	& -13.74 $\pm$0.031	& 0.050 	    $\pm0.075$	& -0.811  $\pm0.063$  &-0.052  $\pm0.276$ \\
RXJ105312.5+573425 & 20 	& 1.20 	& -13.78 $\pm$0.039	&-0.512  	    $\pm0.079$	& -0.507  $\pm0.015$  &-0.567  $\pm0.386$ \\
RXJ105312.4+572507 & 21 	& 0.96 	& -14.22  $\pm$0.067	& 0.588  	    $\pm0.161$	& -0.932  $\pm0.107$  & 0.560  $\pm0.605$ \\
RXJ105316.8+573552 & $^\star$22 & 1.20  & -12.60 $\pm$0.009	& 0.209             $\pm0.022$	& -0.648  $\pm0.021$  &-0.401  $\pm0.053$ \\ 
RXJ105322.2+572852 & 23 	& 2.30 	& -14.13 $\pm$0.062	& 0.539  	    $\pm0.181$	& -0.535  $\pm0.118$  &-0.743  $\pm0.333$ \\
RXJ105324.7+572819 & 24 	& 1.51 	& -14.19  $\pm$0.070	& 0.636  	    $\pm0.197$	& -0.850  $\pm0.110$  & 0.224  $\pm0.473$ \\
RXJ105329.2+572104 & 25 	& 1.14 	& -13.91  $\pm$0.053	& 0.138  	    $\pm0.119$	& -0.896  $\pm0.104$  &-0.305  $\pm1.171$ \\
RXJ105331.8+572454 & 26 	& 1.95 	& -13.18  $\pm$0.017	& 0.223  	    $\pm0.042$	& -0.705  $\pm0.037$  &-0.344  $\pm0.116$ \\
RXJ105335.1+572542 & 27 	& 0.78 	& -12.77  $\pm$0.010	& 0.228  	    $\pm0.025$	& -0.713  $\pm0.022$  &-0.475  $\pm0.069$ \\
RXJ105339.7+573105 & 28 	& 0.58 	& -12.86  $\pm$0.012	& 0.016 	    $\pm0.028$	& -0.757  $\pm0.025$  &-0.739  $\pm0.092$ \\
RXJ105344.9+572841 & 29 	& 1.81 	& -13.44  $\pm$0.025	& 0.236  	    $\pm0.065$	& -0.538  $\pm0.060$  &-0.593  $\pm0.132$ \\
RXJ105350.3+572710 & 30 	& 1.70 	& -13.47 $\pm$0.028	& 0.396  	    $\pm0.071$	& -0.658  $\pm0.058$  &-0.218  $\pm0.163$ \\
RXJ105358.5+572925 & 31 	& 1.84 	& -14.10 $\pm$0.081       &-0.125  	    $\pm0.187$	& -0.657  $\pm0.215$  &-0.010  $\pm0.584$ \\
RXJ105421.1+572545 & $^\star$32 & 0.20 	& -12.19 $\pm$0.007       & 0.711  	    $\pm0.015$	& -0.497  $\pm0.018$  &-0.478  $\pm0.034$ \\

\end{tabular}
\end{center}
$^\star$ Radio Loud \\
$^1$ Assuming a power-law model with $\Gamma=2$  \\
\end{table*}

\begin{figure}
 \centering
 \includegraphics[angle=0,width=8cm,height=5cm]{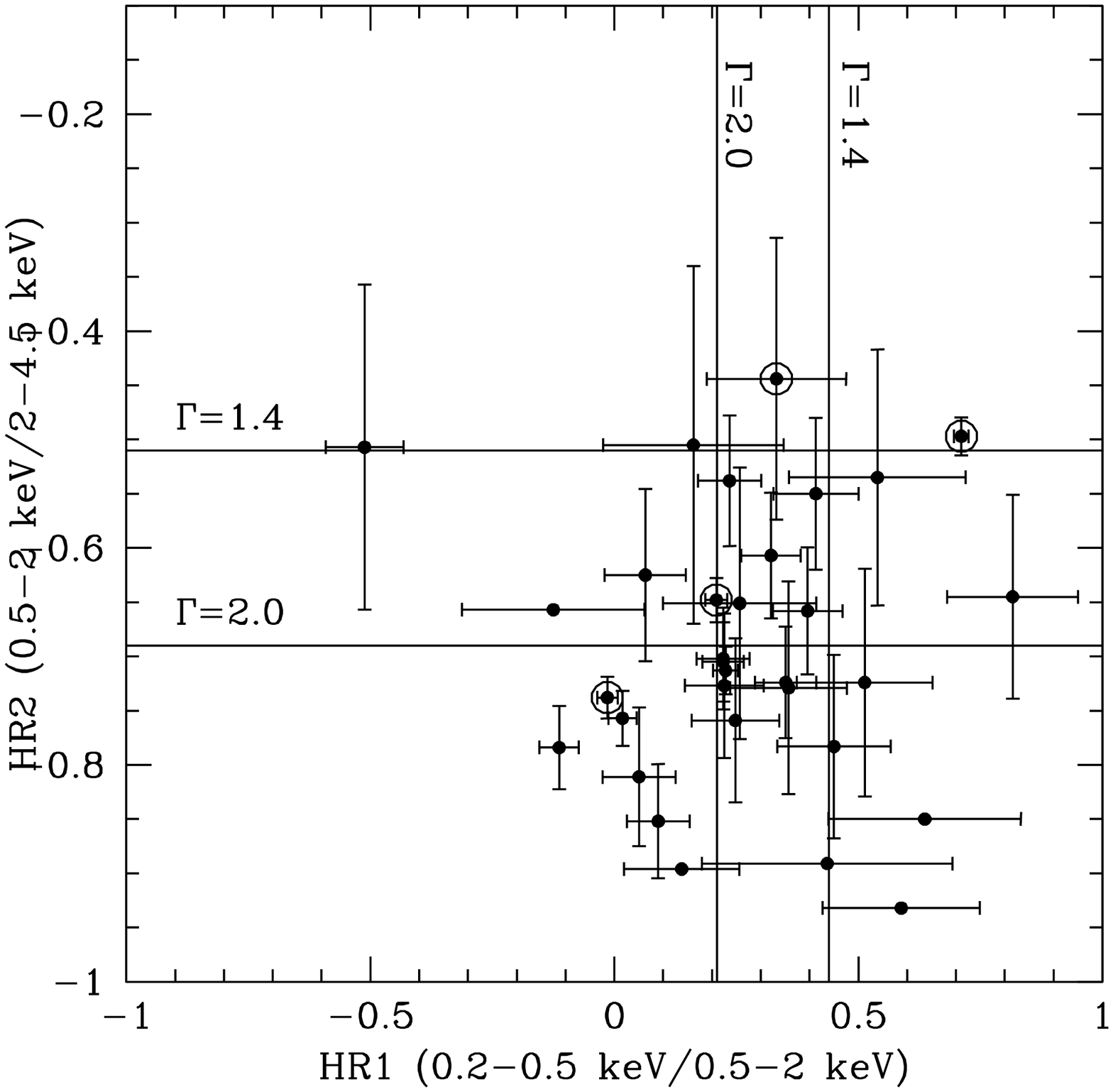}
 \caption{The HR2 (0.5-4.5 keV) as a function of 
 the HR1 (0.2-2 keV) hardness ratios for 
 radio-quiet (solid circles) and radio-loud (open circles) QSOs.  
 The solid lines 
 denote power-law spectra with spectral indices 
 of $\Gamma=2$ and $\Gamma=1.4$  absorbed by a column density of 
 $6\times 10^{19}~\rm cm^{-2}$.}
 \label{hr12}
 \end{figure}

\begin{figure}
   \centering
   \includegraphics[angle=0,width=8cm,height=5cm]{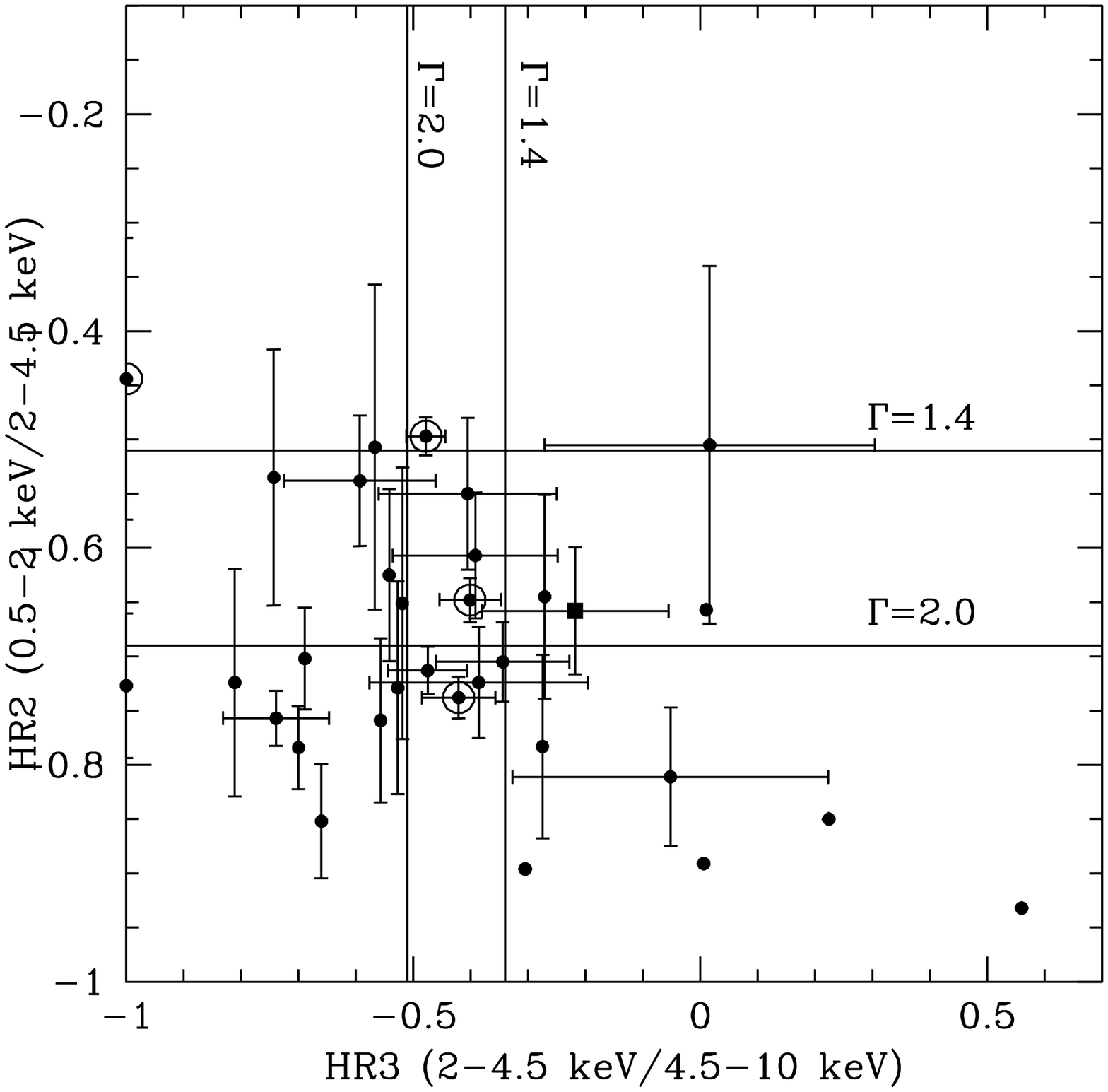}
 \caption{The HR2 (0.5-4.5 keV) as a function of 
 the HR3 (2-8 keV) hardness ratios for 
 radio-quiet (solid circles) and radio-loud (open circles) QSOs. 
 The solid lines denote power-law 
 spectra absorbed by a column density of 
 $6\times 10^{19}~\rm cm^{-2}$.}
   \label{hr23}
   \end{figure}

In Fig.  \ref{hr12} we plot the HR2 versus 
 HR1 for our 32 sources while in Fig. \ref{hr23} 
 we plot the HR2 versus the HR3 hardness ratio.
 For the sake of clarity, we plot the errors only in the cases 
 where we have at least 15 counts in each energy band.  
 In Fig. \ref{hr12} we find    
  several sources with a spectrum 
 harder than $\rm HR1\sim 0.45$ ($\Gamma=1.4$) in the soft band
 (\#1, 2, 21, 23, 24, 32).
 These have softer spectra in the hard (2-8 keV) band 
 (see the HR3 values Table \ref{master}) 
 and therefore absorption is the most likely 
 explanation for the hard HR1 ratios.
 In the case of the source \#32, which is bright enough to allow for 
 spectral fitting, we find that indeed the spectrum is absorbed 
 by an intrinsic column of $\sim 2\times10^{21}$ \cunits.     
 The other five sources are located at higher redshifts 
  so that, if the absorption is intrinsic, 
 the {\it rest-frame} columns are much larger.  
 For example for sources \#1 and 2,  both located 
 at a redshift of $z\approx 3.4$, 
 the observed hardness ratios translate to columns of 
 $\sim 5\times 10^{22}$ and $\sim 3\times 10^{23}$ \cunits 
 respectively, assuming a spectrum with $\Gamma=2.0$.
 This is in agreement with earlier findings by
 Fiore et al (1999), Akiyama et al. (2000)  
 who first presented evidence for the existence of absorbed  
 broad-line QSOs,  using {\it BeppoSAX} 
 and \asca data respectively.  
 In Fig. \ref{hr23}, there are a few sources 
 with a more complex behaviour.
 These show  marginal evidence (given the large error bars in HR3)
 for spectral hardening at high energies ($\rm HR3>-0.3$
 or $\Gamma<1.4$) while their soft spectra  
 are steep (cf Giommi et al. 2000, 
 Barcons et al. 2002). 
 Only one of our sources with a flat hardness 
 ratio at hard energies (HR3) is bright enough (\#30) to allow us 
 to derive a detailed spectral fit (see Fig. \ref{fig_spectra}
 and Table \ref{spectra}).      
 The fit gives a relatively hard spectrum $\Gamma=1.67^{+0.42}_{-0.18}$,
 (albeit with large uncertainty). This spectrum is somewhat  
 steeper, but comparable to that derived from the HR3 hardness ratio 
 (the 68\% upper limit of HR3 corresponds to  $\Gamma\approx1.5$) 
 and in much better agreement with  the
 spectrum derived on the basis of the HR2 hardness ratio 
 (which is consistent with $\Gamma=2$).   
 We therefore believe that there is no conclusive 
 evidence in our data 
 for the presence of a population 
 of {\it soft X-ray selected} QSOs 
 with intrinsically flat spectra at hard energies.

\begin{figure}
   \centering
   \includegraphics[angle=0,width=8cm,height=5cm]{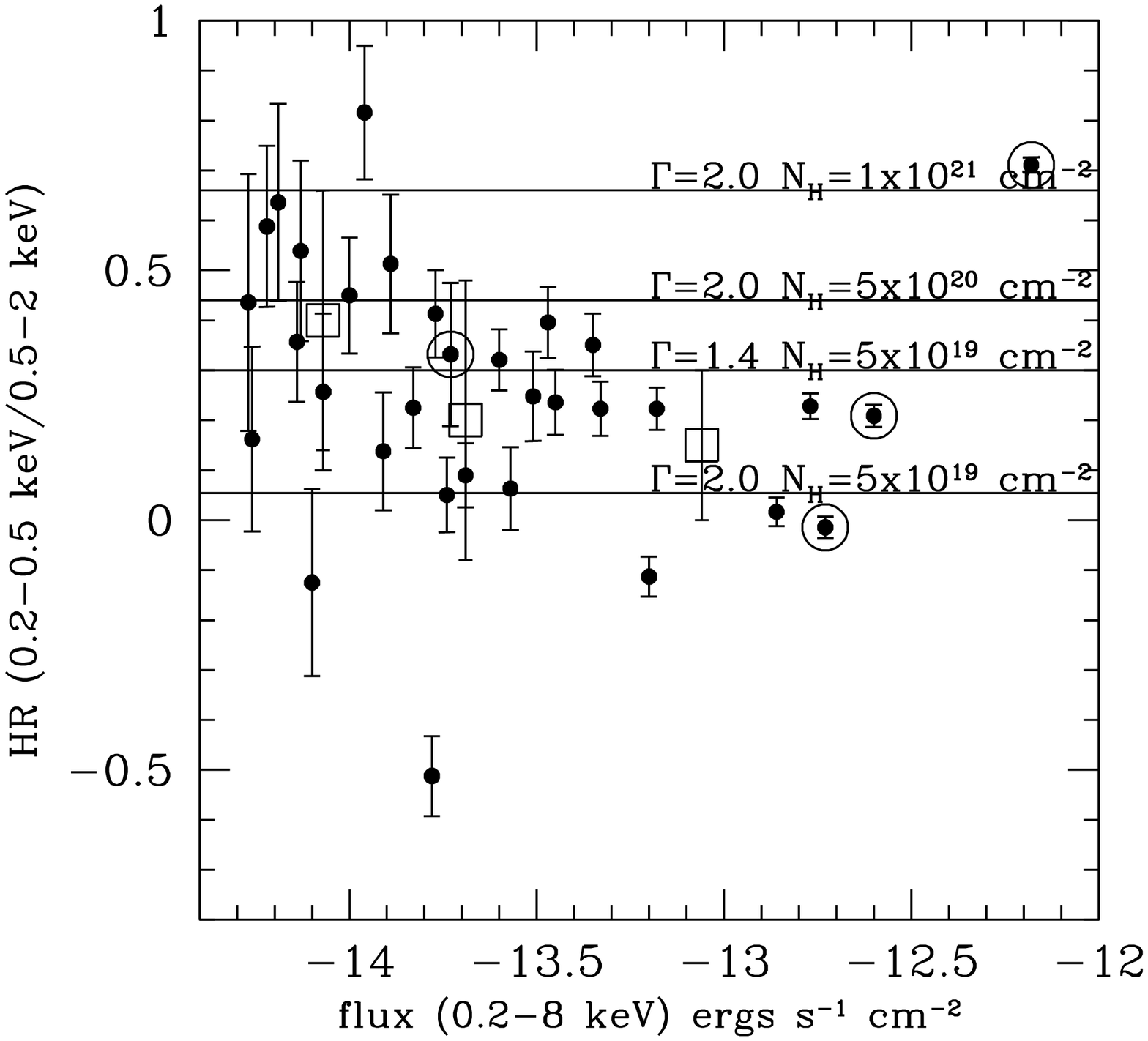}
   \includegraphics[angle=0,width=8cm,height=5cm]{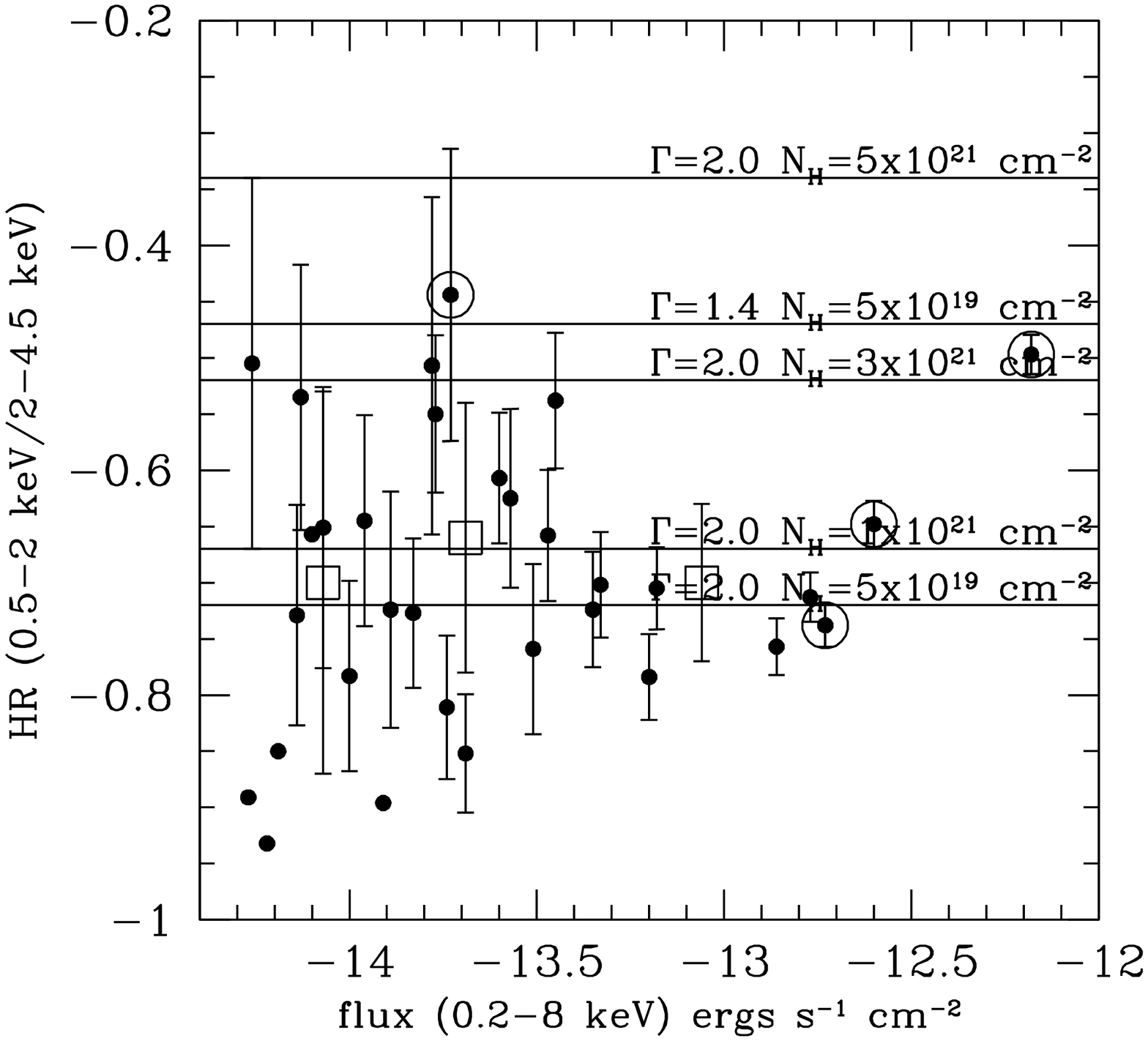}
   \includegraphics[angle=0,width=8cm,height=5cm]{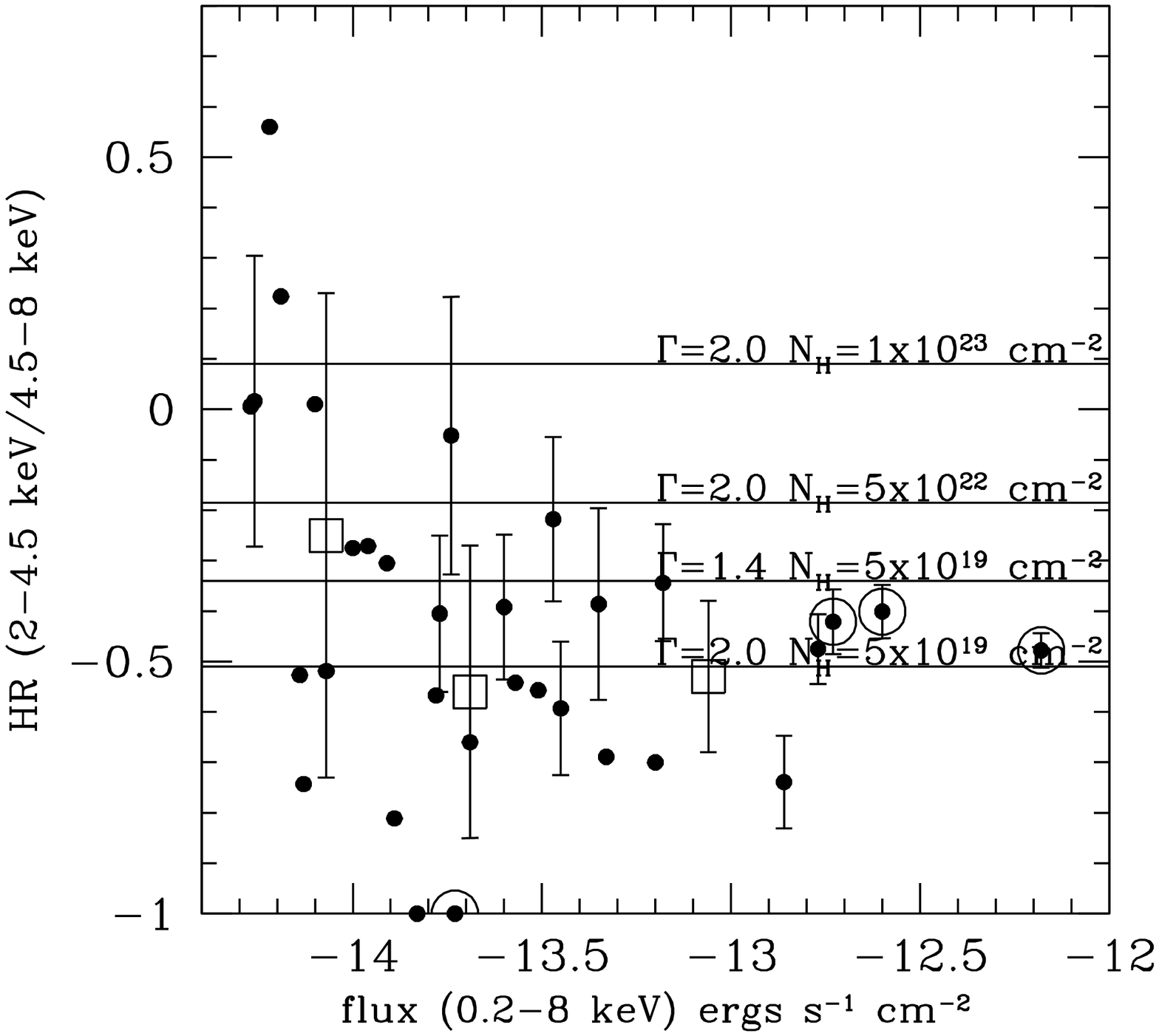}
\caption{ HR values as a function of the  
 0.2-8 keV flux:  HR1 (upper panel), HR2 (middle panel)
  HR3 (lower panel). 
 The solid and open circles denote the radio-quiet and radio-loud QSOs
 respectively. Errors are plotted only 
 for sources having at least 15 counts in each band.  
 The open boxes represent the mean HR values}
   \label{Fig_flux}
   \end{figure}

\begin{figure}
   \centering
  \rotatebox{0}{\includegraphics[width=8cm,height=5cm]{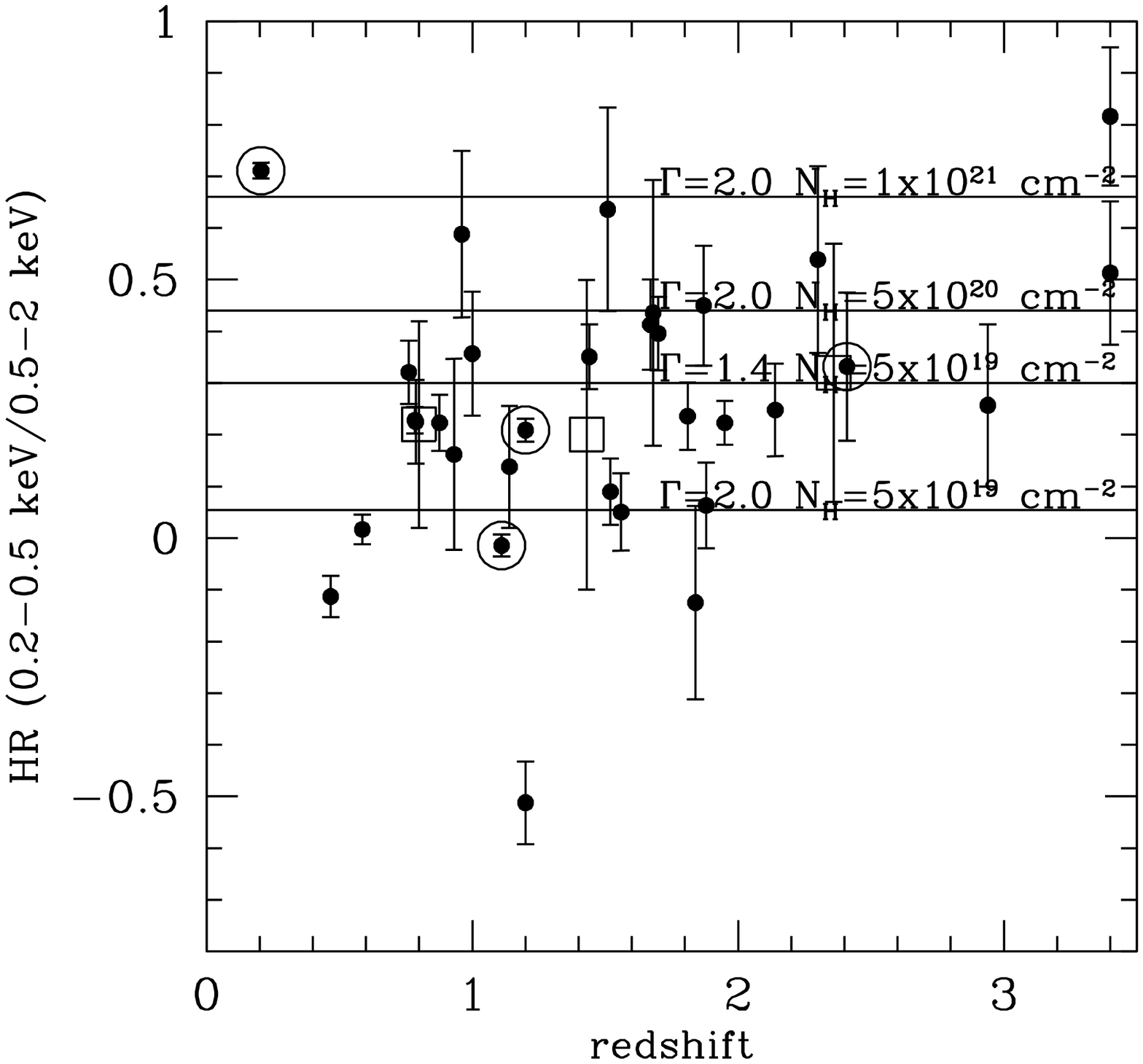}}
  \rotatebox{0}{\includegraphics[width=8cm,height=5cm]{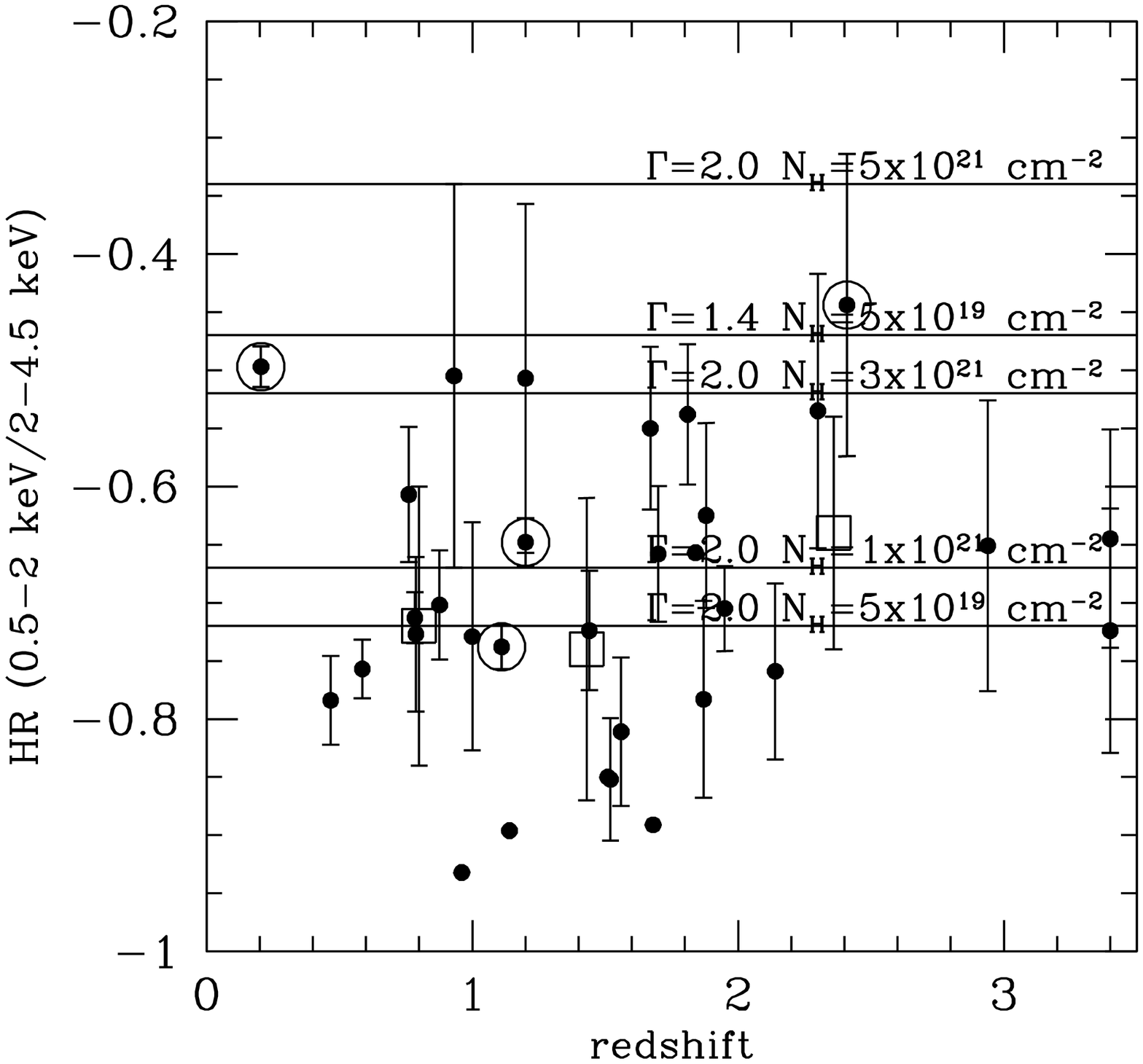}}
  \rotatebox{0}{\includegraphics[width=8cm,height=5cm]{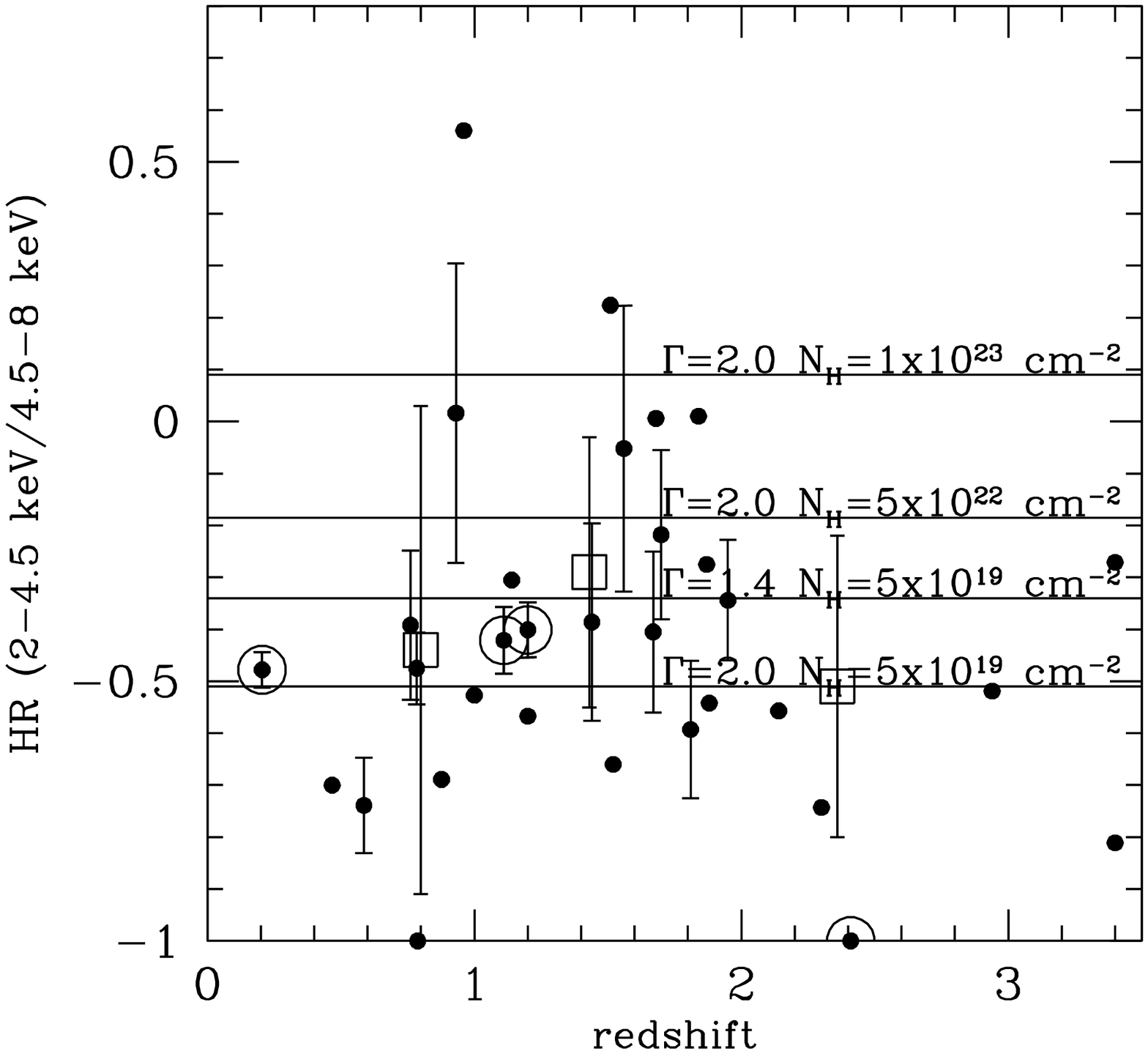}}
 \caption{ HR values of type I QSOs as a function of 
 redshift: HR1 values (upper), HR2 (middle), 
  HR3 (lower panel). 
 The solid and open circles denote the radio-quiet and radio-loud QSOs
 respectively. Errors are plotted only 
 for sources having at least 15 counts in each band.  
 The open boxes represent the mean value} 
   \label{Fig_z}
   \end{figure}

Next, we investigate whether there 
is a correlation between the hardness ratio and the 
X-ray flux. Hasinger et al (1993) and  Della Ceca et al. (1999),  
presented evidence for such a relation, 
 in the sense that the photon index becomes harder with 
 decreasing flux, using \asca and \rosat data. 
 However, their samples contain 
all X-ray sources, not just QSOs in contrast to our sample.     
We divide the data into three flux bins and we  
calculate the mean value of HR. The mean is calculated 
 by averaging the individual hardness ratios in each flux bin
 (no weights for the individual HR errors are applied).
 The advantage of estimating the mean HR value 
 in the way above -as compared to adding up  
 the counts in each bin- is that the much larger errors reflect 
 the intrinsic spectral dispersion of the QSO population.
 We exclude the brighter (and nearest) source of our sample 
 (\# 32) which is a relatively absorbed ($\rm N_H\sim 2\times 10^{21}$) 
 QSO at z=0.2.      
 In Table \ref{table_flux} we present the HR values
 for each flux bin.  In Fig. \ref{Fig_flux} we plot the 
 individual HR values (HR1, HR2, HR3) as a function of flux
 together with the mean HR values for the three flux bins.
 Each bin  contains roughly the 
 same number of sources while the flux corresponds to the 
 middle of the bin.
 There is no statistically significant 
 correlation of any of the HR with flux. 
 Moreover, we explore any possible dependence of the HR on redshift. 
 We derive the mean values of HR in three redshift bins
 (Table \ref{table_z} and Fig. \ref{Fig_z}). 
 We choose the groups so that each one  
 contains roughly the same number of sources. 
 We find no dependence of any of the HR on redshift.
 The lack of relation between HR1 and redshift suggests that 
 the intrinsic column density is not a function of redshift 
 {\it or}  luminosity (as there is a strong correlation between 
 redshift and luminosity). 
  The lack of correlation between 
 HR1 and redshift also comes in contradiction to the results 
 of Schartel et al. (1996) and Blair et al. (2000). 
 However, the {\it number}  statistics of our sample 
 are still  very limited ($\sim$ 10 sources per redshift bin) 
 to allow us to draw definitive conclusions.   
 Finally, the fact that HR3 does not get harder with 
 increasing redshift 
 suggests that the strengths of any reflection components
 at high energies ($>$10 keV) must be small.

\begin{table*}
\begin{center}
\caption{ Mean HR values for three flux bins}
\label{table_flux}
\begin{tabular}{ccccc}
 {\small mean flux (0.2-8 keV)} & $^1$No & HR1 & HR2 &  HR3    \\
$ \rm ergs~s^{-1}~cm^{-2}$      &      &     &     &         \\
\hline 
8.70$\times10^{-14}$ & 9& 0.15 $\pm$ 0.15 & -0.70 $\pm$ 0.07 & -0.53 $\pm$ 0.15 \\     
2.04$\times10^{-14}$ &11& 0.20 $\pm$ 0.28 & -0.66 $\pm$ 0.12 & -0.56 $\pm$ 0.29 \\     
0.85$\times10^{-14}$ &11& 0.40 $\pm$ 0.26 & -0.70 $\pm$ 0.17 & -0.25 $\pm$ 0.48 \\     
\end{tabular}
\end{center}
$^1$ Number of stacked AGN

\end{table*}

\begin{table}
\begin{center}
\caption{ Mean HR values for three redshift bins}
\label{table_z}
\begin{tabular}{ccccc}
mean z & $^1$No & HR1 & HR2 &  HR3    \\
\hline 
0.80 & 9 & 0.22$\pm$ 0.20 & --0.72$\pm$ 0.12 & --0.44$\pm$ 0.47 \\     
1.43 &11 & 0.20$\pm$ 0.30 & --0.74$\pm$ 0.13 & --0.29$\pm$ 0.26 \\     
2.36 &11 & 0.32$\pm$ 0.25 & --0.64$\pm$ 0.10 & --0.51$\pm$ 0.29 \\     

\end{tabular}
\end{center}
$^1$ Number of QSOs in the bin 
\end{table}

\begin{figure}
   \centering
   \includegraphics[angle=0,width=8cm,height=5cm]{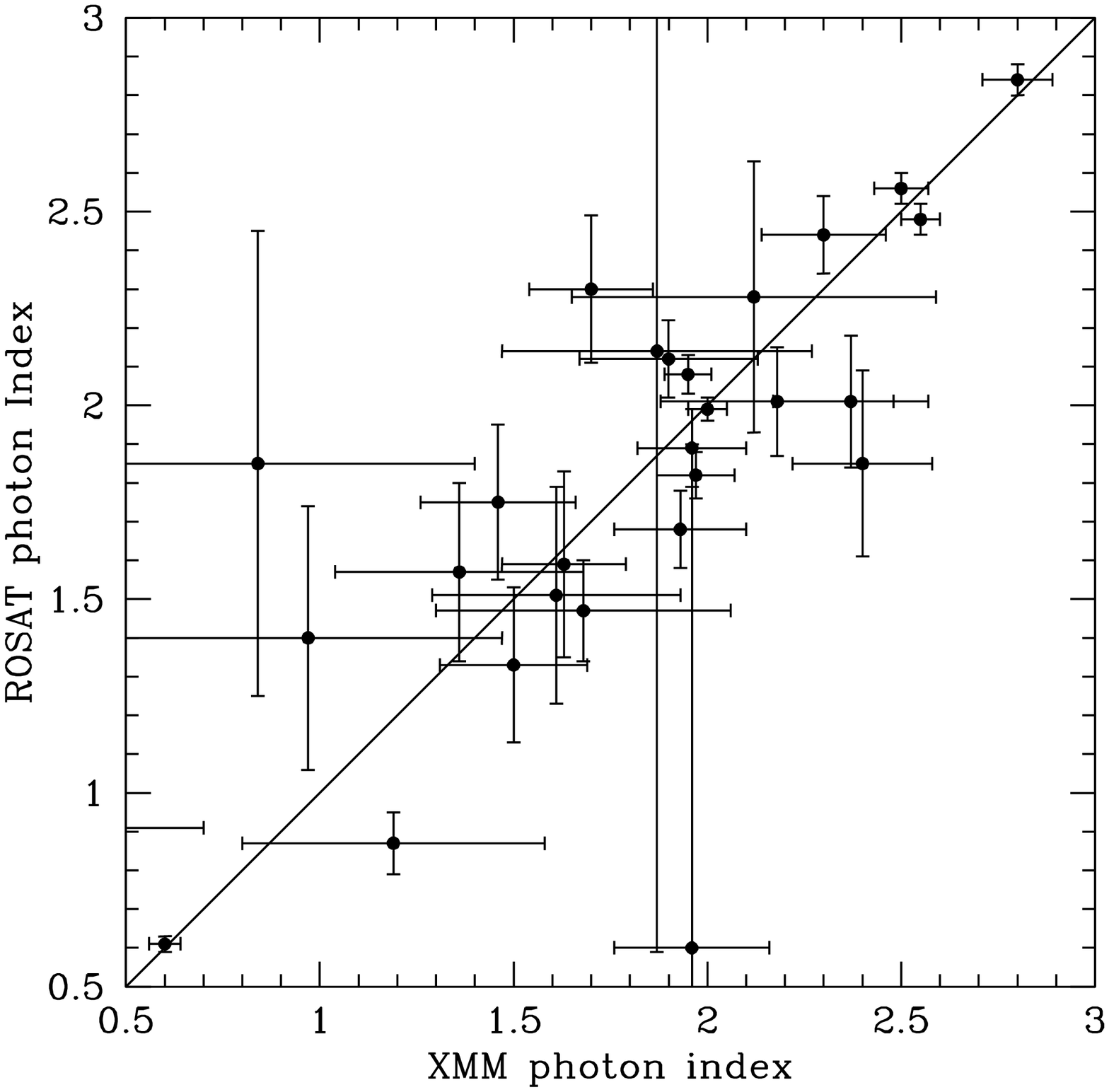}
\caption{ The \rosat versus the \xmm spectrum. The 
 \rosat photon index is derived from the 0.1-2 keV hardness ratio 
 while the \xmm photon index from the 0.2-2 keV hardness ratio}
   \label{onetoone}
   \end{figure}

We also compare the obtained  HR1 values with that presented 
by Lehmann et al (2001) using \rosat 
data. Note that Lehmann et al. (2001)  have calculated  the HR using data 
in slightly different energy bands,  namely 
0.1-0.4 keV and 0.4-2 keV. 
 In order to convert the \rosat and \xmm HR to  photon index we  
 assume a power-law spectrum  of $\Gamma=2$ absorbed by the 
 Galactic  column density $6\times10^{19}~\rm cm^{-2}$. 
 In  Fig. \ref{onetoone} we plot
 the \xmm  photon index 
 versus \rosat  photon index for the 11  bright type-1 QSOs.
 There is very good agreement between the \rosat 
 and \xmm HR1 values in most cases.

\subsection{The average QSO spectrum}

\begin{figure}
   \centering
\rotatebox{270}{\includegraphics[width=6.0cm]{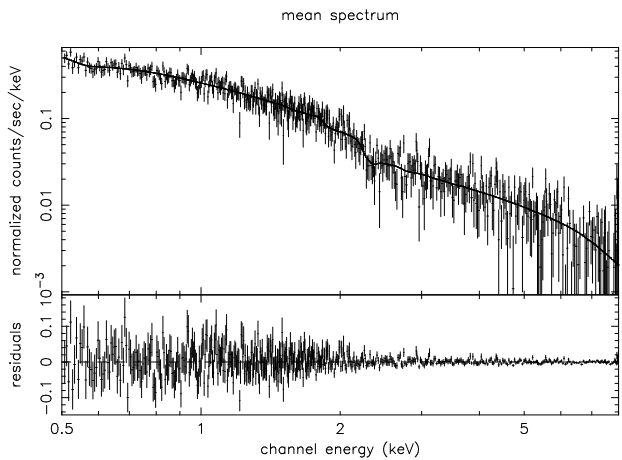}} 
 \caption{ The average spectrum  of the 32 QSOs 
 plotted with the best-fit  power-law model and the 
 $\chi^2$ residuals.}
   \label{32spectrum}
   \end{figure}

 We finally derive the average QSO spectrum in the 0.5-8 keV band
 by co-adding the photons from all 32 QSOs, in the observer's frame. 
 The vignetting and PSF corrections were applied by 
 creating 32 auxiliary files with the {\sc sas} task {\sc arfgen}
 and then co-adding them using the {\sc addarf} task of {\sc ftools}.   
 The spectral fit to a power-law model yields  $\Gamma=1.86^{+0.02}_{-0.02}$
 with column density $\rm N_H\approx 0^{+1}\times 10^{20}$ 
 \cunits ($\chi^2=569.6/524$).
 Although the fit is very good, we tried to fit separately 
 the hard (2-8 keV) and soft (0.5-2 keV) energies in order 
 to check whether there is any hint for a more complex model.  
 We find very good agreement between 
 the two power-laws with $\Gamma_{soft}=1.88^{+0.05}_{-0.04}$ and 
 $\Gamma_{hard}=1.94^{+0.07}_{-0.12}$.
 The average spectrum of 32 QSOs is shown in Fig. \ref{32spectrum};  
 there we also plot the best fit model and the $\chi^2$ residuals.
 The derived mean 'typical' QSO 
 spectrum is significantly steeper than the spectrum of the 
 X-ray background in both soft and hard energies. 
 Indeed, results from all X-ray missions have demonstrated 
  that the spectrum of the X-ray background is 
 $\Gamma=1.4-1.5$ in the 1-10 keV band (Gendreau et al. 1995, 
 Vecchi et al. 1999).  
 As the above average spectrum is dominated by our  
 11 bright sources, we have derived separately the spectrum for the 
 21 faint sources. Again we find a steep  spectrum 
 ($\Gamma=1.94^{+0.10}_{-0.10}$ with $\chi^2=259.4/224$) 
 consistent with the total spectrum.
 Hence soft X-ray 
selected, broad-line AGN with obscured  spectra do not 
contribute significantly to the total QSO flux in this field.   
 Page (1998) also find that the average   
 \asca spectrum of the soft X-ray selected QSOs 
 from the RIXOS sample is steep ($\Gamma\sim1.8$).  
 These results come in apparent contradiction 
 to the findings of Pappa et al. (2001) and 
 Barcons et al. (2002) who find spectral hardening in the average 
 QSO spectrum. However, we note that our sample 
 contains only soft X-ray selected QSOs in contrast 
 to the surveys of Pappa et al. (2001) and Barcons et al. (2002). 
 Soft X-ray selected samples preferentially select soft 
 or unabsorbed sources possibly explaining the apparent discrepancy.

\section{Summary}

 We present the spectral analysis of 32 type-1 (broad-line) QSOs 
 in the Lockman Hole. These faint QSOs 
 ($\rm f_{0.2-8 keV} > 5\times10^{-15}$ \funits)
 contribute an appreciable  fraction ($\sim 50$ \%) 
 of the soft X-ray background. 
 For the 11 bright sources, where good photon 
 statistics are available, we  derive the individual spectra. 
 Most of these 11 bright QSOs present the  
 canonical AGN spectrum ($\Gamma\sim 1.9$) or steeper  
 with little or no significant absorption 
 above the Galactic value.
 These findings are in excellent agreement with 
 previous  \asca results on 
 (predominantly optically selected) QSOs eg Reeves \& Turner (2000).   
 The HR analysis shows  
 evidence for QSOs with large 
 absorbing columns (up to $10^{23}$ \cunits in the QSO's rest--frame).
 This finding is in agreement with previous results by Fiore et al. (1999) 
 and Akiyama et al. (1999). 
 There is no conclusive evidence from our individual 
 spectra or hardness ratios 
 for spectral hardening at high energies. 

 Although there is evidence
  that the spectrum of a few faint QSOs  
 presents large  absorbing columns,  
  the average spectrum of all 32 QSOs 
 has the canonical value of $\Gamma\sim 1.9$.
  This  is much steeper than the spectrum of the 
 X-ray background in the 1-10 keV band. 
 This is also significantly steeper  than the average QSO spectrum found in 
 hard X-ray selected QSO samples (Pappa et al. 2001, Barcons et al. 2002). 
 The accumulation of further, deeper \xmm data 
 as well as future observations with high effective area missions 
 such as {\it Constellation-X} and {\it XEUS} will 
 allow us to shed more light on the spectral properties 
 of moderate to high redshift QSOs.

\begin{acknowledgements}
 We are grateful to the referee Dr. Belinda Wilkes for 
 her numerous corrections and suggestions. 
 This work has been supported by a Greek-Spanish bilateral 
 Scientific Agreement grant under the title 
 ``Search for Obscured AGN with XMM''.
 Partial financial support for XB was provided by the Spanish 
 Ministry of Science and Technology under project 
 AYA2000-1690.  
 This work is based on data obtained 
 from the \xmm public data archive.

\end{acknowledgements}

\end{document}